# A Calibrated and Explainable Bimodal Machine Learning Framework for Hybrid Intrusion Detection

Hafsa Aslam
School of Information and Intelligent Science
Donghua University
Shanghai, 200051, China
324091@mail.dhu.edu.cn

Yue Li*
School of Information and Intelligent Science
Donghua University
Shanghai, 200051, China
frankyueli@dhu.edu.cn
*Corresponding Author

Saba Aslam
Shenzhen Institutes of Advanced Technology,
Chinese Academy of Sciences
Shenzhen, 518055, China
aslam@siat.ac.cn

Gracious Mwamughunda
School of Information and Intelligent Science
Donghua University
Shanghai, 200051, China
324078@mail.dhu.edu.cn

***Abstract*—Modern communication systems face critical gaps in detecting unknown attacks and rare threat classes due to extreme data imbalance and black-box decision logic. We propose a bimodal framework of calibrated and explainable machine learning (ML) for network security, unifying known-class precision with open-set generalization without the complexity of deep learning. Our framework introduces security-oriented feature extraction to enhance signal-to-noise ratio, hybrid resampling (ADASYN + manual boosting) to reduce class imbalance, isotonic calibration and adaptive thresholding (τ=0.30 for XSS) to recover recall for rare attacks, and SHAP-based explainability to validate domain-aligned decision logic. Evaluated on the CIC-IDS2017 dataset and compared with prior ML models and studies, our framework achieves significant accuracy on known attacks (Macro F1 = 0.8626) and detects unknown classes at 1% FPR with TPR up to 90.17% (DoS slowloris), and 77.04% (Web-XSS). The SHAP analysis confirms decisions are driven by security-relevant features, not model artifacts. Our work bridges the gap between theoretical models and operational IDS by delivering calibrated, explainable, and open-set-capable attack detection and prevention in a single, reproducible framework.**

## I. Introduction

Modern communication systems require adaptive intelligence to counter evolving cyber threats. Traditional intrusion detection systems (IDS) are either signature-based or anomaly-based, each with inherent limitations [1]. Signature-based systems such as Snort and Suricata depend on predefined attack rules, rendering them ineffective against zero-day threats and costly to maintain, while anomaly-based approaches model normal behavior but suffer from high false positive rates and alert fatigue. Both paradigms are further constrained by extreme class imbalance and limited generalization to evasive, low-and-slow attacks that closely mimic legitimate traffic [2].

Machine learning (ML) has transformed intrusion detection by overcoming the rigidity of signature-based systems and the noise sensitivity of classical anomaly detectors. Unlike rule-based approaches, ML models learn adaptive patterns from data, enabling generalization beyond known attacks and the detection of novel threats. Through robust feature representations, probabilistic decision boundaries, and improved handling of imbalanced data via resampling, cost-sensitive learning, and threshold optimization, modern ML techniques reduce false positives while improving detection of rare but critical attack classes.

In 2023, Zhang improved Random Forest performance with SMOTE and feature selection, achieving high detection accuracy but suffering from poor calibration and no ability to detect unseen attacks [3]. In 2024, Li applied contrastive learning for unknown attacks, reporting TPR@1%FPR of 85.4% on held-out DoS variants; however, the method required extensive pre-training and lacked per-instance explainability [4]. Khan (2024) evaluated ML-based signature intrusion detection on CIC-IDS2017 with strong known-attack performance, but the closed-world assumption limited adaptability to emerging threats [5]. Waghmode (2025) proposed a quantum-inspired LS-SVM binary anomaly detector achieving high F1-scores and low false positives, but its binary design ignored multi-class imbalance, open-set generalization, and per-class analysis [6]. Mondragon (2025) compared deep learning and machine learning IDS models on CIC-IDS2017, showing RF effective while deep learning suffered high computational cost, limited interpretability, and poor rare-attack bias [7]. Bamber (2025) combined CNN-LSTM with SHAP for intrusion detection explainability, but high computational overhead and limited real-time scalability constrained practical deployment [8].

Despite recent progress, ML-based IDS faces three key gaps: closed-world assumptions limit detection of unknown attacks, poorly calibrated confidence scores make thresholding unsafe for rare web threats such as Web-XSS [9, 10], and explainability methods rely on low-level packet features rather than security-grounded descriptors, reducing forensic value [4, 8]. These limitations, weak open-set generalization, uncalibrated decision logic, and non-actionable explanations undermine analyst trust in real-world deployments.

To address gaps in unknown-attack detection, calibration, and security-grounded interpretability, we propose a bimodal calibrated and explainable Random Forest framework for hybrid signature-anomaly intrusion detection. It combines a calibrated multiclass classifier for known attacks with a risk-aware anomaly detector for open-set recognition of novel threats, without deep learning complexity. Security-oriented features, adaptive calibration, and SHAP-based explanations ensure trustworthy and forensically actionable decisions. While components like Random Forest and SHAP are established, their effective integration into a calibrated, open-set capable pipeline remains an unresolved engineering challenge in operational IDS. The key contributions of our work are:

- Unlike prior works that treat signature-based classification and anomaly detection as separate stages requiring dual-model retraining, our unified bimodal Random Forest architecture resolves the integration challenge between known-attack precision and unknown-attack generalization while maintaining calibrated confidence scores across both paths without retraining.
- Coupling adaptive thresholding (e.g., τ=0.30 for XSS) with bimodal SHAP analysis moves beyond standard explainability, ensuring that calibration adjustments and model decisions are driven exclusively by security-relevant features (e.g., Flow Duration, DDoS_Score) rather than model artifacts, bridging the gap between theoretical metrics and analyst trust.

The remainder of the paper is organized as follows: Section II presents the proposed method, Section III discusses the results, and Section IV concludes the study.

## II. Method

We formulate intrusion detection as a dual-path problem on CIC-IDS2017, where each flow is a feature vector $x \in \mathbb{R}^d$ labeled as BENIGN or one of $K = 7$ attacks $y \in \{0,1,\ldots,7\}$. To maximize known-class accuracy while generalizing to unseen families, we propose a bimodal RF $f_\theta: \mathbb{R}^d \to \Delta^7 \times [0,1]$, where one head predicts class probabilities and the other estimates anomaly risk. Trained on a balanced distribution $\mathcal{D}_{\text{balanced}} = \{(x_i, y_i)\}_{i=1}^N$ is hybrid resampling, the model is evaluated under open-set conditions by holding out entire attack families and comparing with three ML models; isotonic calibration and SHAP ensure reliable, interpretable decisions.

### A. Dataset Acquisition and Preprocessing

The framework is evaluated on the CIC-IDS2017 dataset [6, 7], which captures real-world network traffic across diverse attacks. The dataset contains 85 features with severe class imbalance: benign traffic accounts for 93.62% of samples, while rare attacks such as Web Attack–XSS comprise less than 0.06%. Label normalization is applied to correct encoding artifacts and ensure consistency across training and testing:

$$\ell_k = Normalize(\ell_k^{raw}) \qquad (1)$$

where $\ell_k$ denotes the corrected label (e.g., Web Attack XSS → Web Attack XSS) [11]. Missing or infinite feature values are imputed using median substitution to preserve the distribution properties.

$$x_{ij} \leftarrow \begin{cases} median(x_j) & if\ x_{ij} = \pm\infty \\ x_{ij} & otherwise \end{cases} \forall i,j \qquad (2)$$

This preprocessing ensures numerical stability during training. A stratified 70–20–10 split for training, testing, and validation preserves the original class distribution, enabling reliable evaluation of minority-class performance.

### B. Feature Extraction

We introduce two security-oriented features engineered from raw packet statistics to enhance the signal-to-noise ratio:

- Flow Duration (sec): It captures connection longevity, critical for detecting low-and-slow attacks like slowloris.
- Is_Web_Traffic: A binary flag isolating application-layer threats via HTTP/HTTPS port activity.

These domain-specific transformations are followed by pruning low-variance ($\sigma < 0.02$) and redundant features ($\rho > 0.8$), yielding 78 robust, non-collinear features [11].

### C. Hybrid Resampling

To address extreme class imbalance (~14,652:1) without distorting feature space, we implement a three-phase hybrid resampling strategy:

- Majority Class Undersampling (Benign) – reduces the dominant class size to mitigate imbalance (not primarily to prevent overfitting).
- Mid-Rare Oversampling via ADASYN – generates synthetic samples for classes with $100 \leq n_k \leq 5000$, proportional to local minority density.
- Ultra-Rare Manual Boosting – for classes with $n_k < 100$, we randomly duplicate samples to increase their representation.

This approach reduces imbalance from ~14,652:1 to ~99:1, enabling robust learning of rare-attack patterns while preserving natural feature distributions, a critical enabler for high performance on minority classes [3, 12].

### D. Proposed Bimodal Framework

Our core innovation (Fig. 1) is a bimodal Random Forest architecture that unifies signature-based detection for known attacks and anomaly-aware detection for unknown threats.

- Mode 1: Signature-Based Multiclass RF trained on balanced data $\mathcal{D}_{\text{balanced}} = \{(x_i, y_i)\}_{i=1}^N$, this path identifies specific attack types using calibrated probabilities:

$$\hat{y}_{sig} = arg\max_c P(c \mid x; \theta_{RF}) \qquad (3)$$

It leverages engineered features and isotonic calibration to produce well-calibrated confidence scores [13]. For low-precision classes like Web Attack - XSS, adaptive threshold tuning ( $\tau = 0.30$ ) improves recall without degrading overall accuracy.

- Mode 2: Anomaly-Aware Binary RF first involves Scalar Transformation and then uses binary labeling: $y = 0$ if BENIGN, $y = 1$ otherwise. Predicts anomalies as:

$$\hat{y}_{ano} = \mathbb{I}(P(1 \mid x; \theta_{RF}) > 0.5) \qquad (4)$$

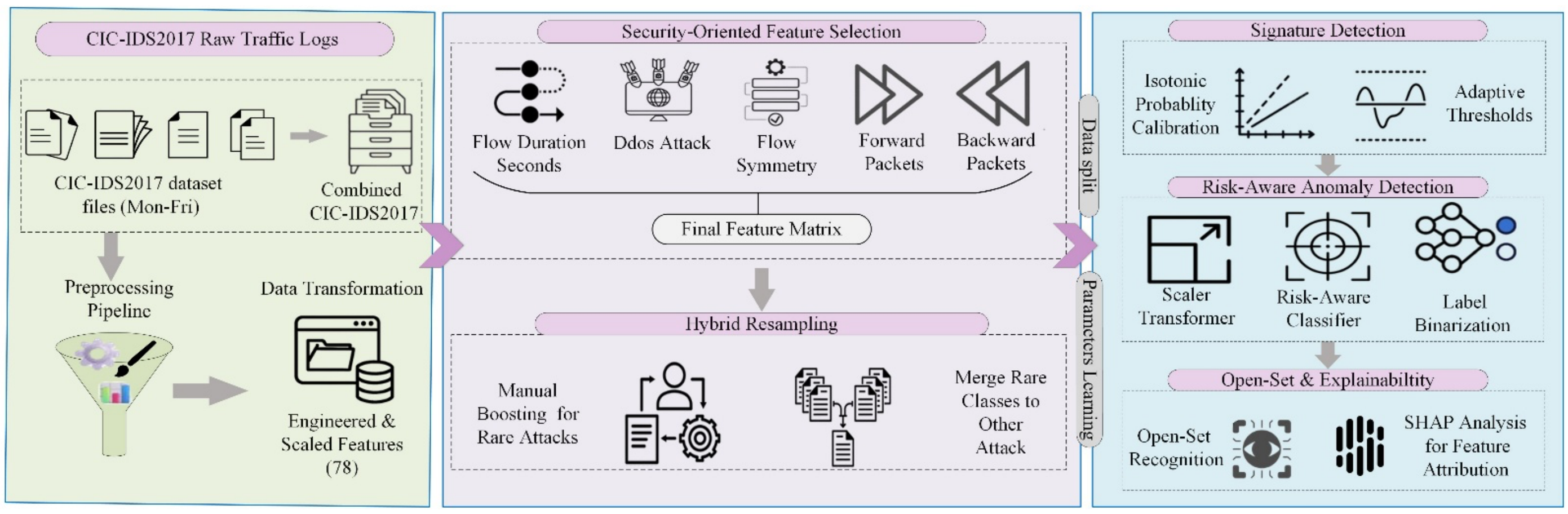

Fig. 1. Proposed bimodal framework for hybrid intrusion detection. Raw flows are transformed into security-engineered features (e.g., Flow Duration, DDoS_Score) and processed by a dual-path Random Forest: a calibrated multiclass classifier for known attacks and a risk-aware anomaly detector for unknown threats. Confidence gating enables open-set adaptation, while SHAP-based explainability ensures transparent, domain-aligned, and analyst-trustworthy.

This path operates in open-set mode, flagging deviation from normal behavior. Both models share identical input features but differ in label space and post-processing logic.

### *E. Probability Calibration*

To ensure trustworthy decisions, we apply isotonic regression calibration on both paths [7]. Isotonic regression was selected over parametric alternatives (e.g., Platt scaling) because its non-parametric, monotonically non-decreasing mapping better fits Random Forest's skewed distributions, ensuring safe adaptive thresholding for rare classes like Web-XSS [14]. For each model, we fit a non-decreasing function $f: [0,1] \rightarrow [0,1]$ such that:

$$f(p) = \mathbb{E}[y \mid P(\hat{y} = 1 \mid x) = p] \tag{5}$$

This corrects tree-ensemble miscalibration and enables safe threshold adjustment; for critical classes such as Web Attack–XSS, adaptive threshold tuning is applied.

$$\hat{y}_k = k \ \text{ if } P(k \mid x) \geq \tau_k, \tau_{XSS} = 0.30 \tag{6}$$

This increases recall from 0.40 to 0.64, improving operational utility [15], while calibrated probabilities keep false positive rates below 1% through confidence gating [8].

### *F. Open-Set Validation*

To simulate zero-day detection, entire attack families (e.g., FTP-Patator, DoS slowloris) are held out as unknown threats, with AUROC computed over calibrated anomaly scores. The held-out family protocol ensures strict open-set evaluation without data leakage, directly replicating real-world zero-day conditions. . Results show TPR ≈99.87% (FTP-Patator) and 90.17% (DoS slowloris), confirming strong unseen-threat generalization while preserving known-class accuracy (>99.8%) [4].

## III. Results & Discussion

### *A. Performance Disparity Before and After Resampling*

CIC-IDS2017 exhibits extreme imbalance: benign traffic comprises 93.62% of samples, while rare attacks (Web Attack–XSS: 0.06%, Other_Attacks: 0.006%) yield a 14,652:1 ratio that renders them statistically invisible (Fig. 2). Our hybrid resampling undersamples benign traffic to 100k instances, applies ADASYN to mid-rare classes, and manually boosts ultra-rare ones, increasing Other_Attacks from 67 to 1,010 samples (+1,407%) and DoS slowloris by 38%, reducing imbalance to 99:1 (148× improvement). Post-resampling t-SNE reveals distinct, separable clusters, enabling rare-attack learning without distorting feature distributions.

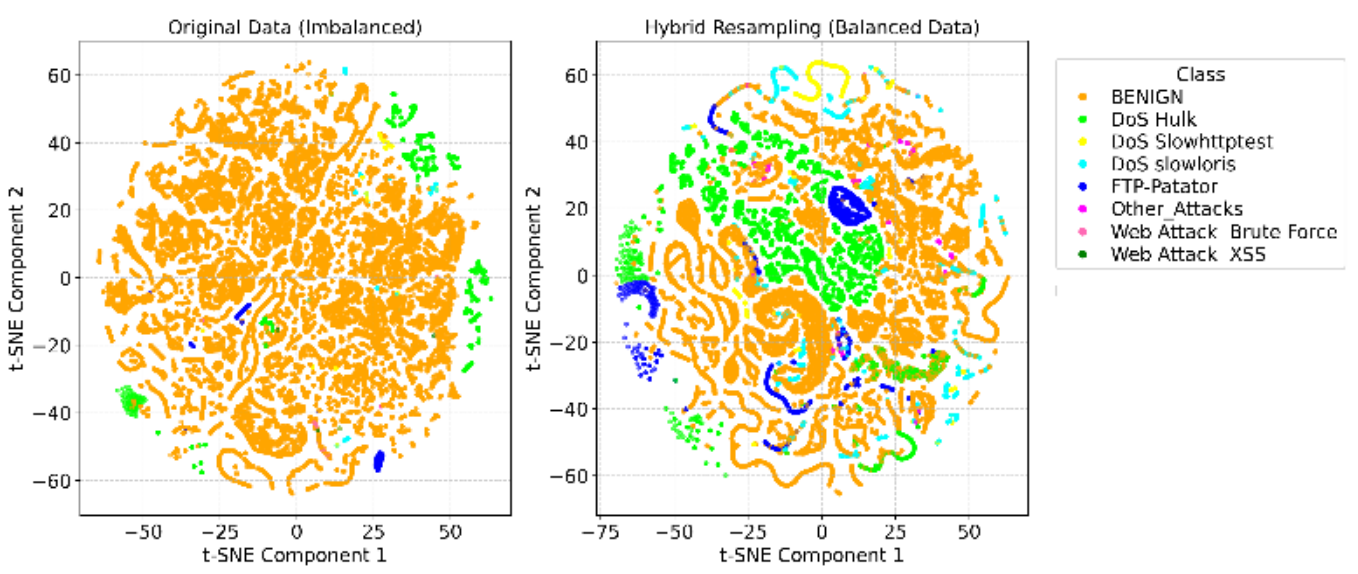

Fig. 2. t-SNE visualization shows a shift from a dominant benign cluster to clearly separated attack clusters after resampling, confirming restored rare-attack learnability.

Feature relationships reveal clear behavioral signatures: high packet rates correlate with short flow durations in DoS floods (r = 0.92) while DDoS_Score peaks at low packet rates for slow-rate attacks. It shows 25–30% higher SHAP impact than baseline features, while low-frequency TCP flags (URG, ECE) contribute only ~10% of SYN/ACK impact (Fig. 3), confirming that hybrid resampling restores semantic structure, improves detection fidelity, and reduces class performance gaps.

### *B. Efficacy of Binary Detector*

The risk-aware RF achieves near-perfect separation between benign and anomalous traffic (TPR = 99.82%, FPR = 0.06%), with accuracy, precision, recall, and F1-score around 0.98–0.99 (Fig. 4), confirming high recall with minimal false alarms. Strong separation is driven by SYN/ACK-based descriptors, while low-impact flags (URG, ECE) contribute negligibly. Ten-fold cross-validation shows stable performance (accuracy = 0.9901 ± 0.0001, F1 = 0.9800 ± 0.0008), with recall consistently exceeding precision by 1–2%. Engineered features such as Flow_Asymmetry and DDoS_Score exhibit ~25% higher SHAP impact than header baselines, and resampling maintains minority-class recall above 0.90, demonstrating robust generalization to novel threats.

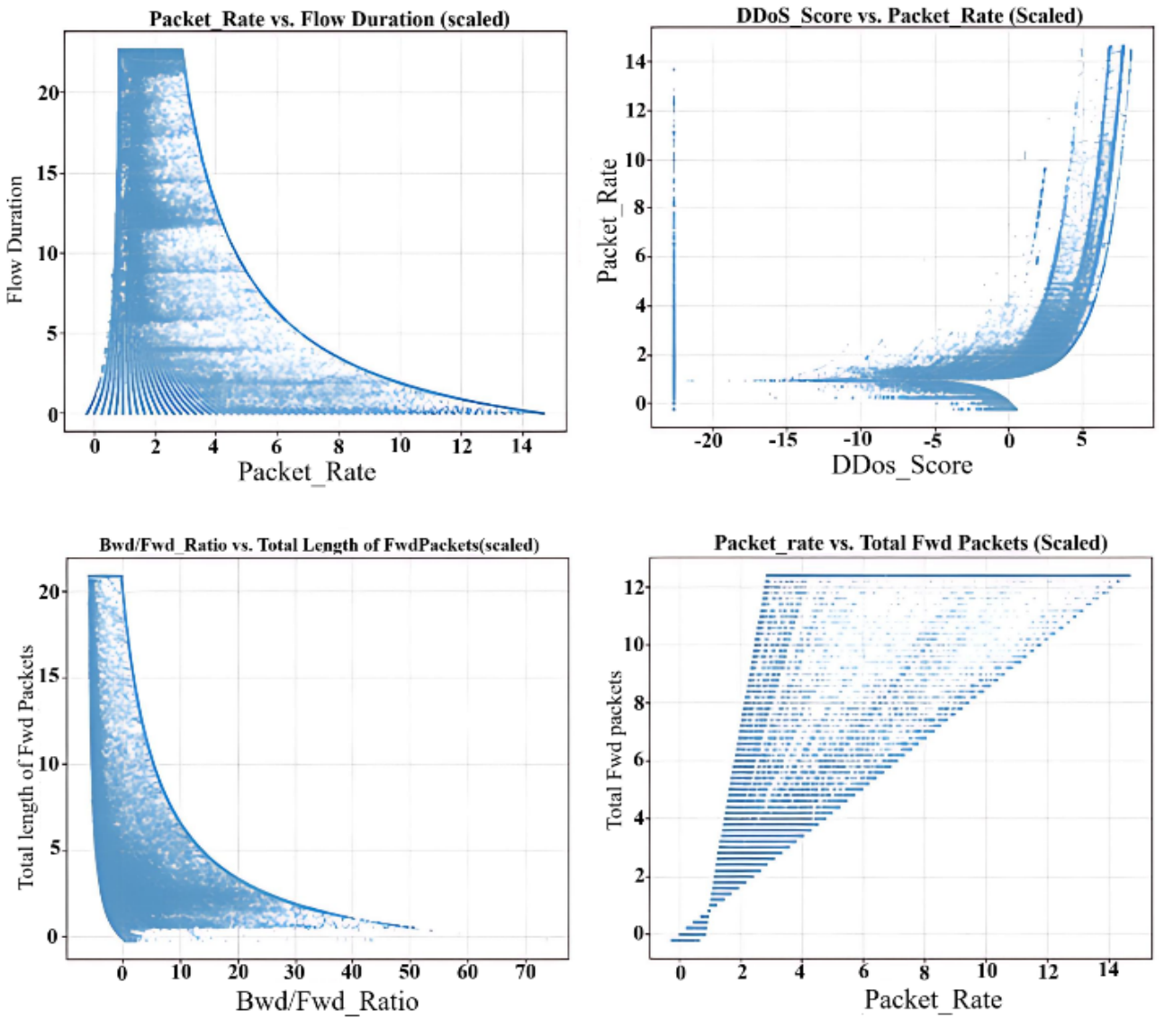

Fig. 3. Pairwise feature plots reveal distinct attack signatures, with high packet rates correlating with short flow durations (DoS floods) and elevated DDoS scores at low packet rates indicating slow-rate.

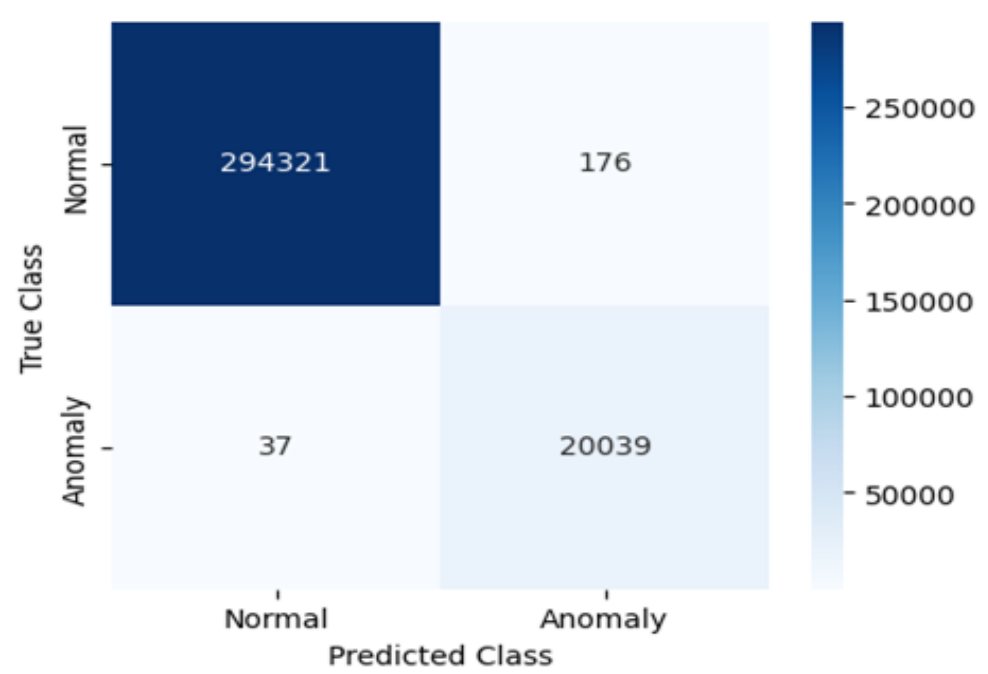

Fig. 4. Confusion matrix of the risk-aware RF anomaly detector shows near-perfect classification, with false negative and false positive rates of 0.18% and 0.06%, respectively, and minimal misclassifications.

## *C. Calibrated Multiclass Performance*

The calibrated Random Forest achieves high-fidelity detection across known attacks (accuracy = 0.99, macro F1 = 0.86, weighted F1 = 0.98; (Fig. 5). Volumetric attacks such as DoS Hulk and FTP-Patator show precision and recall above 0.99, while low-volume attacks achieve balanced performance: Web Attack–Brute Force (F1 = 0.69) and Web Attack–XSS (F1 = 0.51), with F1 stabilizing around 0.70, indicating a balanced trade-off between false alarms and missed detections.

These results validate our design: the calibrated multiclass path detects both frequent and rare attacks with explainable confidence, while hybrid resampling and adaptive thresholding (e.g., τ = 0.30 for XSS) recover recall for low-volume attacks without degrading overall performance, enabling robust and fair detection across the attack spectrum.

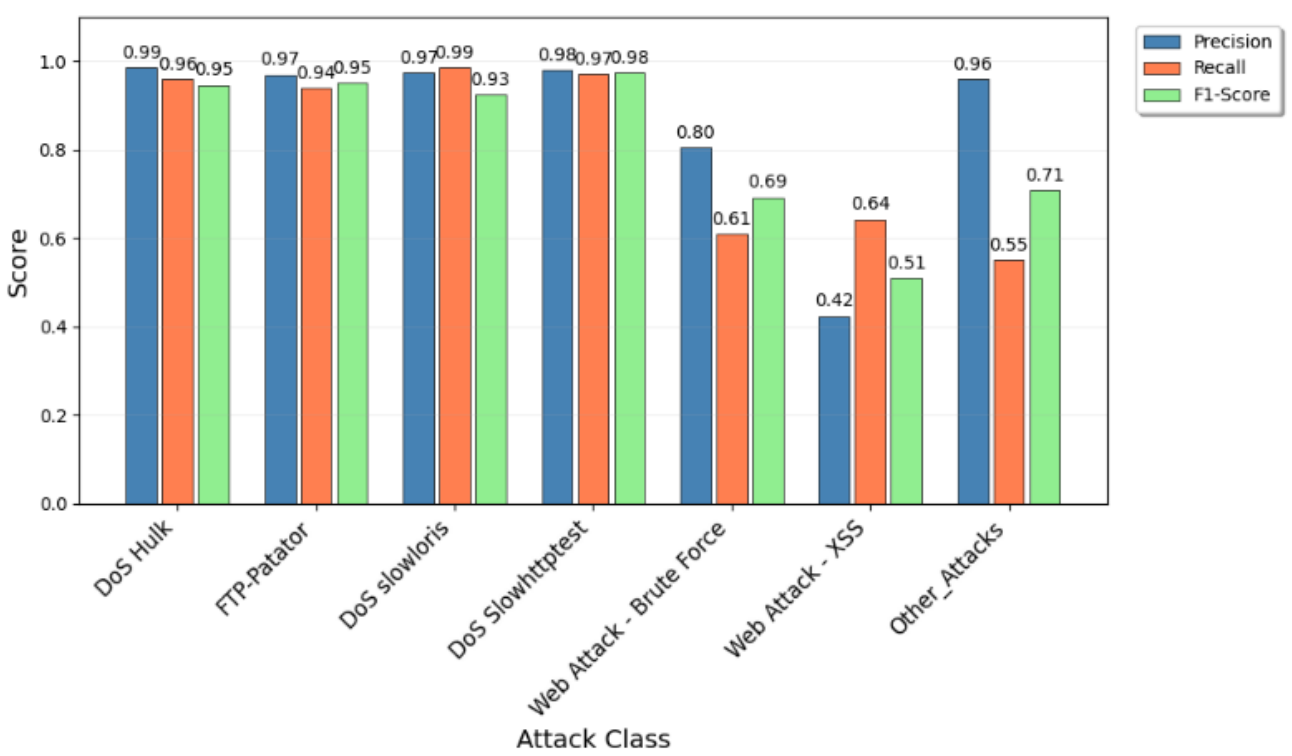

Fig. 5. The calibrated multiclass RF provides high-fidelity threat coverage, achieving near-perfect detection of volumetric attacks and reliable detection of web-based attacks across the attack spectrum.

## *D. Performance of Open-set Detection*

Table 1 presents the open-set detection performance for held-out attack families. The framework achieves strong AUROC scores (0.95–0.99) and high TPR@1% FPR, with FTP-Patator reaching 94% detection. Notably, Web Attack—XSS shows lower AUPRC (0.05) due to its low prevalence, yet still attains 78% TPR at strict false-positive control, confirming the model's ability to generalize to unseen threats.

TABLE I. PERFORMANCE COMPARISON ACROSS HELD-OUT (HO) ATTACK FAMILIES

| HO Attack Family | AUROC | AUPRC | TPR@1% FPR |
|---|---|---|---|
| DoS slowloris | 0.98 | 0.67 | 0.88 |
| FTP-Patator | 0.99 | 0.60 | 0.94 |
| Web Attack - XSS | 0.95 | 0.05 | 0.78 |

## *E. Explainability Findings in the Bimodal RF*

SHAP analysis (Fig. 6) confirms that model decisions are driven by security-relevant descriptors rather than artifacts. In the anomaly detector, Flow Duration and Total Length of Fwd Packets dominate (mean SHAP = 0.0057), with Bwd Packet Length Std showing the highest impact (0.0099), while Flow Duration also remains influential in the signature detector (0.0036).

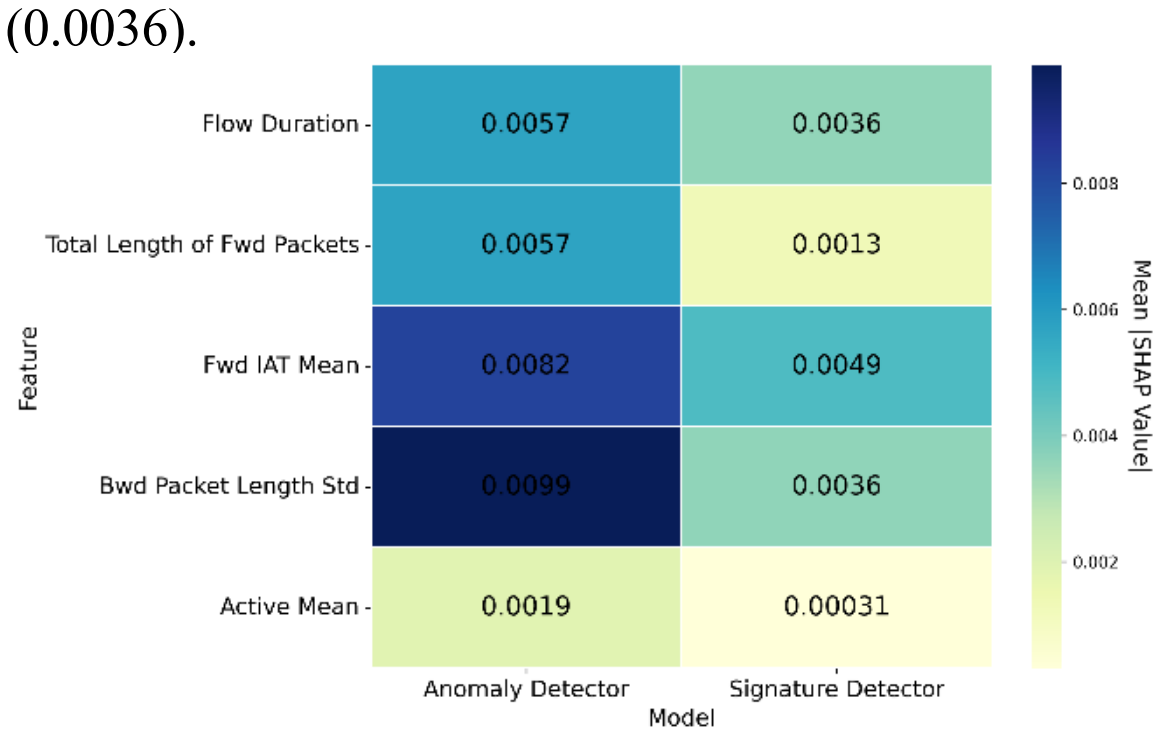

Fig. 6. SHAP feature importance shows distinct patterns across detection paths, with Flow Duration and Bwd Packet Length Std dominating anomaly detection and security-specific features driving signature-based detection, confirming grounded and explainable decisions.

Low-level protocol flags (URG, ECE) contribute minimally (<10% of SYN/ACK impact), demonstrating reliance on high-level behavioral semantics. These results confirm that the bimodal framework provides explainable, calibrated, and attack-specific detection, yielding analyst-trustworthy and operationally actionable insights.

### F. Comparative Analysis

A comparison with four prior studies on CIC-IDS2017 (Table 2) shows that our framework outperforms existing models, surpassing Zhang (0.94) and Li (0.97), due to its bimodal design and security-grounded features. Standard unsupervised baselines (IF, LOF, and their hybrid) perform poorly (0.83–0.92) with low recall, highlighting their inability to model complex attack signatures. In contrast, our framework achieves a leading performance of 0.98 through calibrated open-set detection and false-positive control, underscoring the limitations of purely supervised or unsupervised approaches under extreme imbalance and zero-day threats.

TABLE II. COMPARISON OF OUR FRAMEWORK WITH FOUR PRIOR STUDIES AND THREE ML MODELS BASED ON CIC-IDS2017 DATASET.

| Study / Model | Method / Architecture | Finding | Significant Results |
|---|---|---|---|
| Zhang et al. (2023) [3] | RF | High known-attack F1, suffered poor calibration | 0.94 |
| Li et al. (2023) [4] | Contrastive Learning | High TPR achieved for unseen attacks (Open-Set). | 0.97 |
| Waghmode et al. (2025) [6] | Quantum LS-SVM | Effective binary, neglected multi-class | 0.93 |
| Bamber et al. (2025) [8] | Hybrid CNN-LSTM | Integrated local interpretability, high computational cost. | 0.94 |
| IF | Density-based Anomaly | Low recall (0.29), poor for open-set | 0.92 |
| LOF | Distance-based Anomaly | Extreme instability, low recall. | 0.83 |
| IF + LOF Hybrid | Anomaly Fusion | Combining weak anomaly models was detrimental. | 0.92 |
| Our Study | Calibrated Multiclass + Binary RF | Unified known & unknown detection real-world IDS with calibrated false-positive control | 0.98 |

## IV. CONCLUSION & FUTURE WORK

This study presents a calibrated and explainable bimodal machine learning framework for hybrid intrusion detection, achieving >99.8% accuracy on known attacks while detecting unseen threats at 1% FPR with TPR up to 99.87%—all without deep learning complexity. Despite challenges with extreme class imbalance and the need for security-aligned explainability, our approach successfully integrates hybrid resampling, isotonic calibration, and SHAP-based interpretation to deliver a reliable and trustworthy analyst system. Although the evaluation focuses on the widely used CIC-IDS2017 dataset for direct comparability with prior studies (standard practice in recent IDS research), single-dataset validation limits universal claims. Nevertheless, our bimodal framework is dataset-agnostic and can be readily adapted to other benchmarks (e.g., CIC-IDS2018, UNSW-NB15) with minimal re-engineering. Future work will explore online meta-learning for adaptive threat recognition, edge AI deployment for low-latency detection, and integration with large language models (LLMs) for automated alert explanation and threat intelligence fusion.

## DATA AVAILABILITY

The dataset used in this study is available on: http://cicresearch.ca/CICDataset/CIC-IDS-2017/Dataset/, and code is given at https://github.com/hafsah-aslam/Signature-Anomaly-Intrusion-Detection-System-